\documentclass[pra,twocolumn,amsfonts,amssymb,amsmath,floatfix,floats,a4paper]{revtex4-1}

\usepackage{graphicx}
\begin{document}

\title{Analysis of counterfactuality of  counterfactual communication protocols }

\author{Lev Vaidman}
\affiliation{Raymond and Beverly Sackler School of Physics and Astronomy\\
 Tel-Aviv University, Tel-Aviv 69978, Israel}

\begin{abstract}
Counterfactual communication, i.e., communication without particle travelling in the transmission channel, is a bizarre quantum effect. Starting from interaction-free measurements  many protocols achieving various tasks from counterfactual cryptogrphy  to counterfactual transfer of quantum states were proposed and implemented in experiments. However, the meaning of conterfactuality in various protocols remained a  controversial topic. A simple error-free  counterfactual protocol is proposed. This protocol and its modification are used as a test bed for analysis of  meaning of counterfactuality to clarify the counterfactuality status of various counterfactual proposals.
      \end{abstract}
\maketitle



\section{INTRODUCTION}
\label{sec:intro}  

The name ``counterfactual''   was coined by Penrose \cite{Penrose}  for describing quantum interaction-free measurements (IFM) \cite{IFM}. It was applied for  counterfactual cryptography  \cite{Noh} and
 counterfactual computation \cite{Joz,Ho06}, but became more widely known after introducing the term
 counterfactual communication \cite{Salih} where it was stated

 \begin{quote}
  ``We show how in the ideal limit, using a chained version of the Zeno effect\cite{Ho06},
information can be directly exchanged between Alice and Bob with no physical particles traveling between them, thus achieving direct counterfactual communication.''
 \end{quote}

In my understanding, ``in the ideal limit'' is about vanishing probability of the failure of the protocol  in the limit of large number of ideal optical elements,  the probability of  events which are discarded by the rules of the protocol.  If we are ready to consider only successful events, we need not to  apply Zeno effect. The word ``direct'' means that we send a message directly and not by first establishing  a secret key which can be achieved by using devices transmitting bit 1 only \cite{Noh}.
I will consider ``counterfactual communication protocol'' as a  direct communication protocol, i.e. as a protocol  capable to transmit both bit 0 and bit 1.  I will not require small probability of a failure in an attempt of communication.
What I want to analyze in this paper is the meaning of the word ``traveling".

In cases that the quantum state of a particle is not described by a localized wave packet, the standard quantum theory does not tell us where is the particle. What is more relevant for the case of a successful counterfactual protocol when the photon was detected is the question: Where  {\it was} the photon responsible for the transfer of the information? Standard quantum theory has even less to say regarding this question. Apparently, Bohr would say that we should not ask the question: Was, or was not, the photon traveling between Alice and Bob?
I find that we {\it can} make useful claims in discussing this question. The answer is not ``yes'' or ''no''. It is a consideration of  possible meanings of ``traveling'' of a quantum particle and corresponding classification of counterfactuality of various protocols.

\section{The counterfactual communication protocol }
\label{sec:proto}

The communication device is an interferometer, part of which is in Alice's site and part is in Bob's site, see Fig.~1. It is a particular combination of  Mach-Zehnder interferometers (MZIs) tuned  to complete destructive interference of some output ports. Detector $D_1$ is a dark port when the interferometer is free from disturbance. The  interferometer which has a part on Bob's site is tuned to destructive interference toward Alice's site, see Fig.~1a. The additional requirement is that when Bob blocks arm $B$ of the interferometer, detector $D_2$ on Alice's site becomes a dark port (while detector $D_1$ cease to be a dark port). This can be achieved by properly tuned phases when beam splitter $BS_1$ is $3:8$, beamsplitter $BS_2$ is $1:2$, and all other beam splitters are $1:1$.

  \begin{figure} [ht]
   \begin{center}
     \includegraphics[width=8cm]{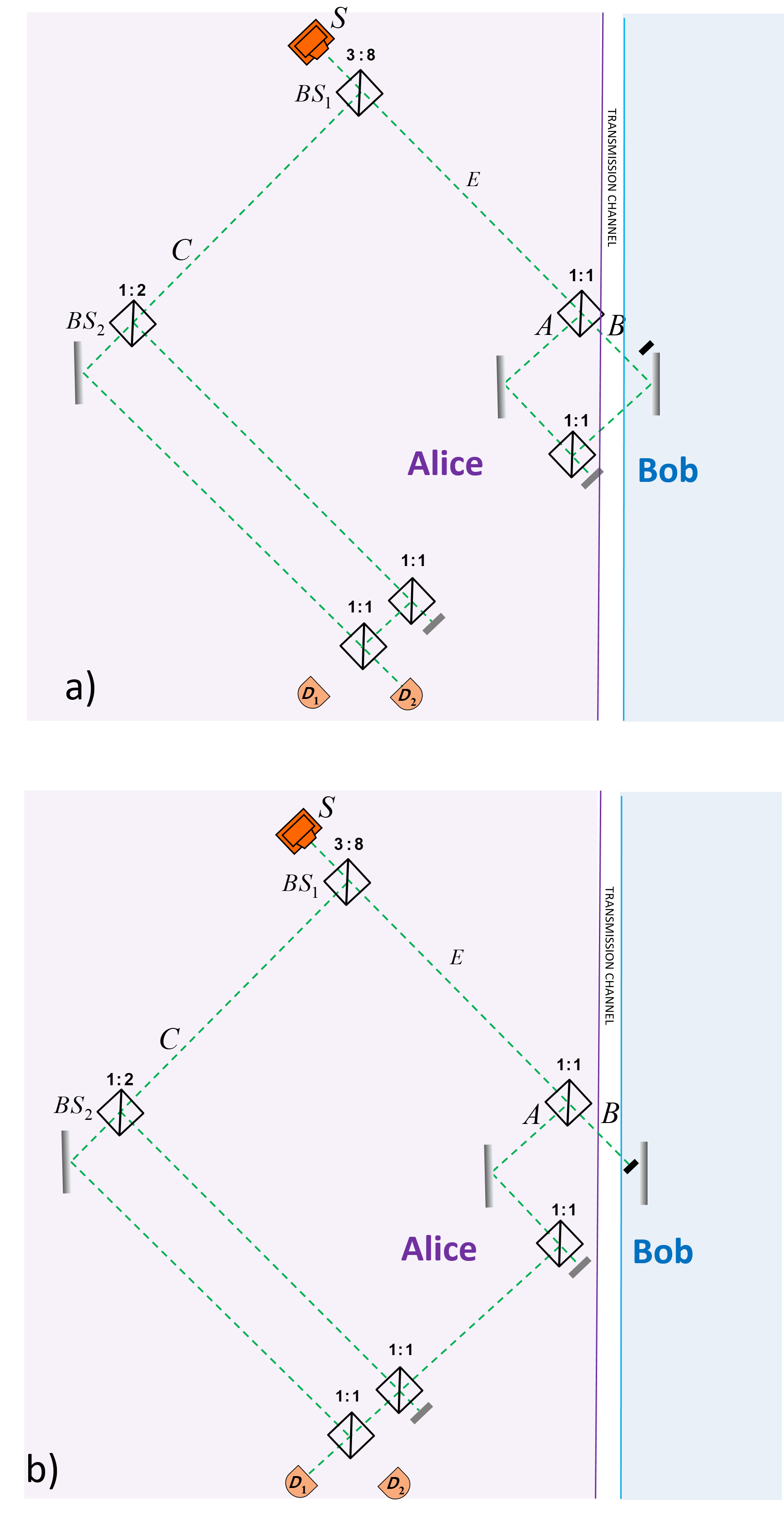}
     \end{center}
   \caption[example]
   { \label{fig:example}
The interferometer with a single photon  source $S$. a) It is tuned such that internal MZIs have destructive interference in the left output ports, and, therefore,  a destructive interference at detector $D_1$. b) It is also tuned such that  when arm $B$ is blocked, there is a destructive interference at detector $D_2$ .}
   \end{figure}

The communication protocol is as follows. At particular agreed time, Bob transmits a value of a bit  to Alice by   blocking the arm $B$ of the interferometer (value 1),   or by leaving it open (value 0).  Alice sends single photons into the interferometer from input port $S$ until she gets a click in detector $D_1$ or $D_2$. The click at $D_1$ tells Alice that the bit value is 1 and the click at $D_2$ that the bit value is 0. If the interferometer made with ideal optical elements and  is perfectly tuned, then there will be no errors in the communication.

Eve, placed between Alice and Bob, has some  efficient options for active attacks for which there are some ways of defence, but  security of this protocol against eavesdropping is not an issue in this analysis. Our question is: ``Whether or not the photon ``traveled'' between Alice and Bob?'' Clearly, some of the photons will travel between Alice and Bob. Those absorbed by the shutter (if present) and those lost after exiting the interferometer towards blocks at Alice's site, but these events are discarded. The protocol is defined for photons detected by Alice's detectors $D_1$ and $D_2$.

Maybe the cleanest way to consider counterfactuality question is to limit ourselves to the ``lucky'' cases of communication with a single sent photon which is detected by Alice on the first run.  Probability for such an event is 2/11, independent on the value of the bit. Current counterfactual protocols, such as \cite{Salih} use many more beam splitters (or recycle them) and apply quantum Zeno effect to make probability of the first run to be successful  close to 1.

Our protocol is considered counterfactual because we can claim, arguing in a classical way, that the  photon did not travel between Alice and Bob.
Our classical physics assumption is that the photon must have a continuous trajectory between the source and the detector. We also make a natural assumption that the trajectory can pass  only through regions where the photon wave does not vanish. We might not be able to know the trajectory, but we assume that it exists. This is an approach pioneered by Wheeler \cite{Wheeler} and recently advocated by Englert et al. \cite{Berge}, see discussion in \cite{PV,PVrep}.

To show counterfactuality we need to consider two cases: bit value 0 when the interferometer is empty, and bit value 1 when the path $B$ is closed. From Fig.~1a and Fig.~1b we see that in both cases there is no wave packet starting at source and reaching Alice's detectors  passing through Bob's site. Classically, for bit 1 we can say that the photon can reach Bob only through path $B$, but if it was there, it must have been absorbed by Bob's shutter. For bit 0 there is no shutter, and photon placed at path $B$ can reach Alice's detectors. However, the only way for a photon to reach path $B$ is through path $E$. Every photon from path $E$ interferes  destructively toward Alice's site, so it also cannot reach Alice's detectors.

\section{Quantum analysis }
\label{sec:quant}

Photon is not a classical object. It is a quantum particle. If the photon wave is a localized wave packet, then it moves like a classical particle on a continuous trajectory, but if the wave packet splits, standard quantum theory does not tell us where the photon goes. A possible answer,  that it was in all places where the wave function  does not vanish, seems inappropriate since we  want to ask where was a pre and {\it post}selected photon, but this approach does not take into account the postselection.

 \begin{figure*} [ht]
   \begin{center}
     \includegraphics[height=16cm]{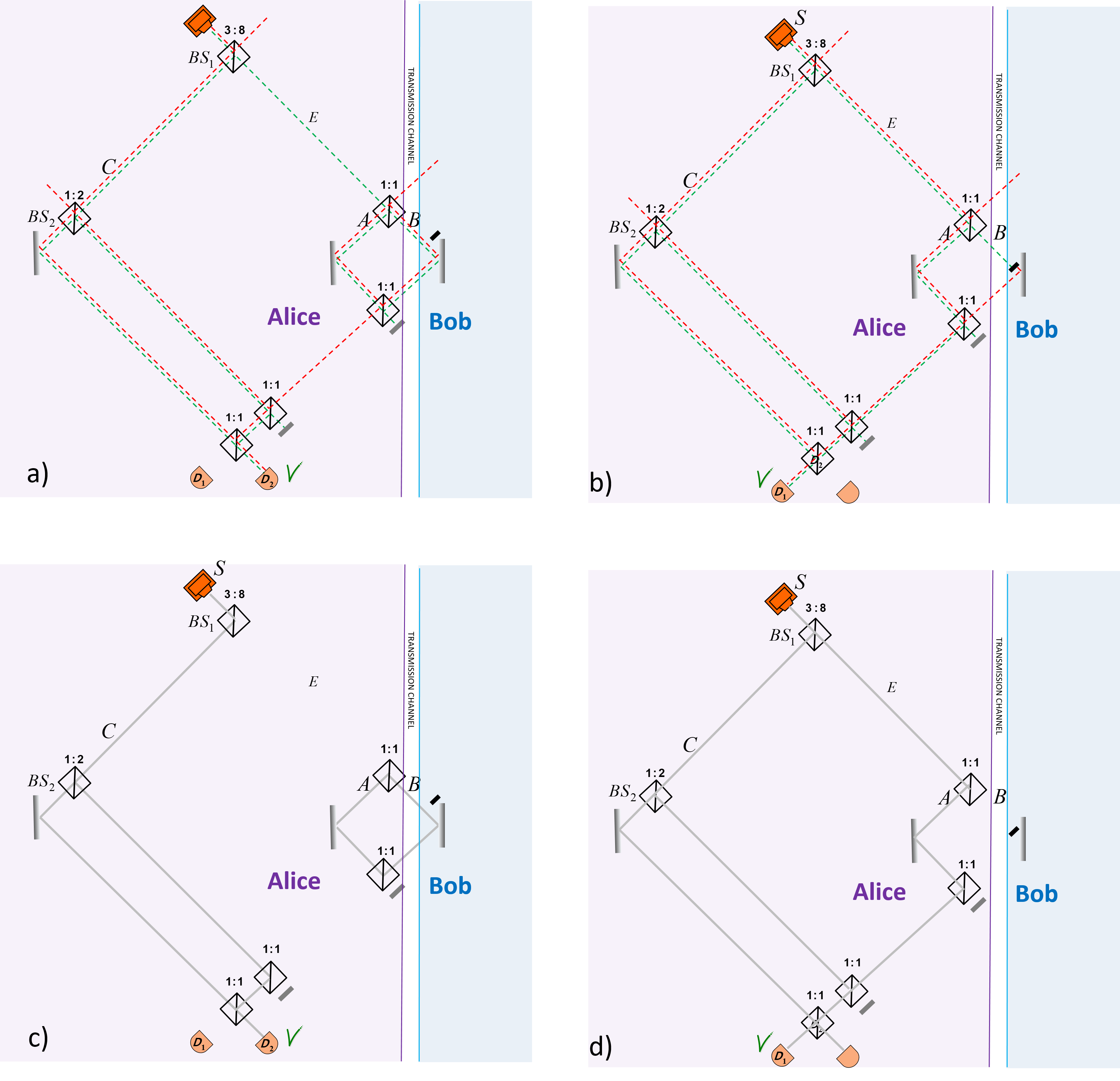}
     \end{center}
   \caption[example]
   { \label{fig:2}
Derivation of the trace left by the photon using two-state vector formalism. a) Forward (dashed green line) and backward (dashed red line) evolving waves of the photon for transfer of bit 0. b) Forward and backward evolving waves of the photon for transfer of bit 1. c) Trace of the photon for transfer of bit 0. d) Trace of the photon for transfer of bit 1. }
   \end{figure*}

Since standard quantum mechanics does not tell us were was a pre and postselected photon we have to consider possible definitions: What does it mean that the photon present or not present in some place? An operational definition of the presence of a classical particle might be the following:

{\it The particle is in a particular location when we know  that the probability of finding it there is 1. The particle is not present in this location when  the probability of finding it there is 0.}

Classically, these are the only options: either the particle is present or not. For quantum particle we might adopt these definitions with understanding that they do not cover all cases. These leads to the following definitions.

{\it The particle was in a particular location if the probability  of the outcome 1 of the measurement of projection operator on this location is 1. The particle is not in this location when  the probability of the outcome 1 of the measurement of projection operator on this location is 0.}

These definitions have  a counterfactual meaning. We make statements about the probability of the results of measurements, if performed, when it is assumed that they were not performed. Quantum measurements change the state of a quantum particle, and we are interested to describe situations when this change was not done. My preferred definition of the presence of a particle is not counterfactual \cite{past}:

{\it The particle was in a particular location if and only if it left a trace there.}

The trace is a change of the quantum state of the environment in this location. It must be of the order of the trace the localized wave packet of  the particle would leave there, the case in which the question of presence of the particle is not controversial. In the interferometer this trace is weak, the quantum state of the environment does not become orthogonal to its state without presence of the photon, but it never vanishes because there are always some local interactions.

The trace definition agrees with counterfactual measurements definition. The easiest way to see this is to use the two-state vector formalism \cite{AV90}. There is a theorem \cite{AV91} that for dichotomic variables such as projection operator, the weak value equal to the eigenvalue if and only if the result of the (counterfactual) strong measurement is obtained with probability 1.
Weak value describes all weak couplings and thus quantifies the trace the particle leaves. Therefore, if we know that the photon to be found with probability 1, it will leave a trace equal to the trace of a localized photon placed there. On the other hand, if we know that the probability to find the photon is 0, then there will be no trace.

The trace definition also covers the case when the probability to find the photon neither 0 nor 1. In this case the trace definition states that the particle is present. Indeed, in this case there will be a weak trace of the order of the trace of a single photon, otherwise, the weak value must be zero and then the probability to find the photon must be zero, contrary to our assumption.

The two-state vector formalism also provides a very simple way to know where the photon leaves a trace. To have a weak trace in a particular location, the weak value of a local operator there must not vanish. Thus, the  requirement is that at this location there is an overlap of the forward and backward evolving states. In Fig. 2. the forward and backward evolving states and the traces are shown for the two cases, bit value 1 and bit value 0. We see that for value 1, the photon does not leave a trace outside Alice's site, but for value 0 there is a trace at the transmission channel between Alice and Bob. We can conclude then that the communication protocol for value 0 is not counterfactual, the photon ``traveled'' between Alice and Bob.

One can argue \cite{Pusey}, that we can define sites of Alice and Bob differently,  such that, for bit value 0 there will be no trace at the new ``transmission channel'' between Alice and Bob, see Fig.~3a. But for this definition of the transmission channel, the photon will be there when the bit value is 1, see Fig.~3b. Whatever definition of the separate sites of Alice and Bob are made, the photon will be in the transmission channel for (at least) one value of the bit. So, if ``counterfactual'' means no particles in some part of the transmission channel, the protocol is not counterfactual.

\begin{figure} [ht]
   \begin{center}
     \includegraphics[width=8cm]{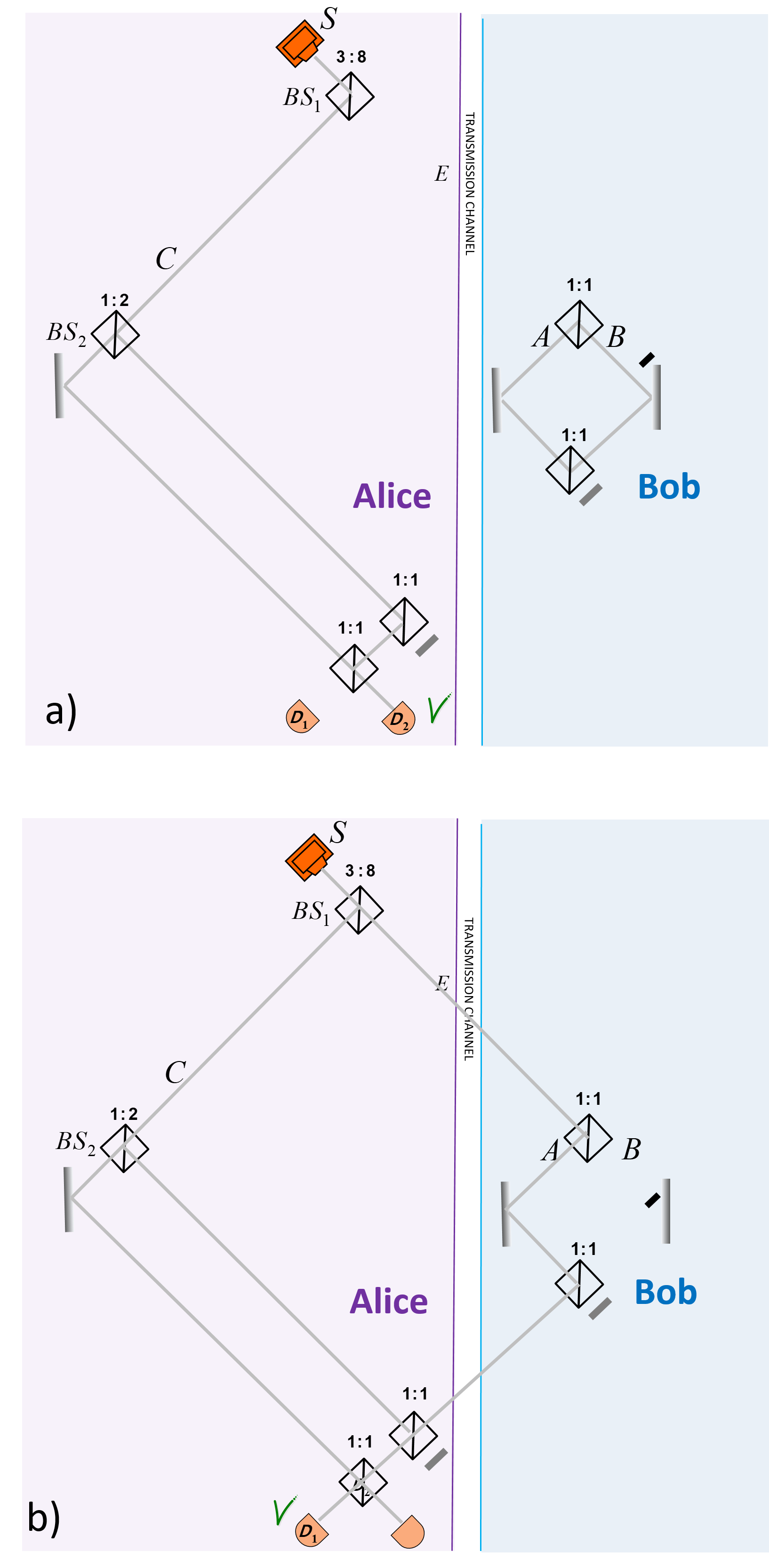}
     \end{center}
   \caption[example]
   { \label{fig:2}
Trace in the transmission channel for modified sites of Alice and Bob. a) No trace in the transmission channel for transfer of bit 0. b) But the trace  in the transmission channel is present for transfer  of bit 1. }
   \end{figure}

In this approach the definition of ``traveling'' between Alice and Bob is following a continuous trajectory between Alice and Bob. In this sense, the photon in the protocol does not travel between Alice and Bob because for any bit value there is a region between them in which the photon leaves no trace. Gisin \cite{Gisin} noted, that if this is the only requirement of counterfactuality, then one can construct a classical ``counterfactual'' protocol which, however, requires a help of a ``middle man'' who sends a  photon to Bob if he does not get a photon from Alice at a particular time.

Denying the idea of continuous trajectories for quantum particles \cite{Danan} leads to defining ``traveling'' of a quantum particle between Alice and Bob as  being in Alice's and Bob's sites. Here again ``being'' defined as leaving a trace. In this sense, the protocol is not counterfactual for bit 0. The ``direct counterfactual communication'' \cite{Salih} and all other published ``counterfactual'' protocols are also not counterfactual in this sense: at least for one bit value the photon leaves a trace both in Alice's and Bob's sites.

I adopt the definition of a counterfactual communication protocol as the one in which the particles left no trace  outside Alice's site. In the next section we will show that such protocol is possible.

\section{Modified protocol }
\label{sec:mod}

The  noncounterfactuality of the ``counterfactual'' proposals is a common feature of numerous proposals \cite{Shen13,Shen14,Li14inv,Guo14ent,guo14,guo15,Li15,chen15,Shen15,chen16,salih16,Guo17,cao17,liu17,zaman18,guo18}. We should exclude indirect counterfactual proposals, based on on key distribution,  which transmit only one bit value  \cite{Noh,zhang13,K-Du,salih14tri}. It led me to conjecture that noncounterfactuality of direct communication is unavoidable property \cite{V07,V14,V14R,V15,V16C,V16R}, but it turned out to be a mistake \cite{AV19}. The modified counterfactual communication protocol, which is counterfactual also according to the trace criterion, is presented in Fig.~4. It is very similar to the original protocol except for replacing the MZI with a mirror at Bob's site by two consecutive MZIs both tuned to destructive interference toward the path continuing inside the large interferometer and readjusting the transmissivity of the beam splitter $BS_1$ to 3:32. According to the new protocol, to transmit the bit 0, Bob should not touch his part of the interferometer, while for bit 1 he has to block {\it two} interferometers, i.e., he has to block  paths $B$ and $B'$. If the interferometer is free,  detector $D_1$ cannot click while detector $D_2$ has probability to detect the photon 2/35. If Bob blocks the two interferometers, detector $D_2$ cannot click while detector $D_1$ has probability to detect the photon 2/35. We, again, consider the ``lucky'' communication in which the first  photon sent by Alice was detected by one of her detectors. (It is possible, with more mirrors and beam splitters, to devise a protocol with higher probability of success of the first run.)

\begin{figure} [ht!]
   \begin{center}
     \includegraphics[width=7cm]{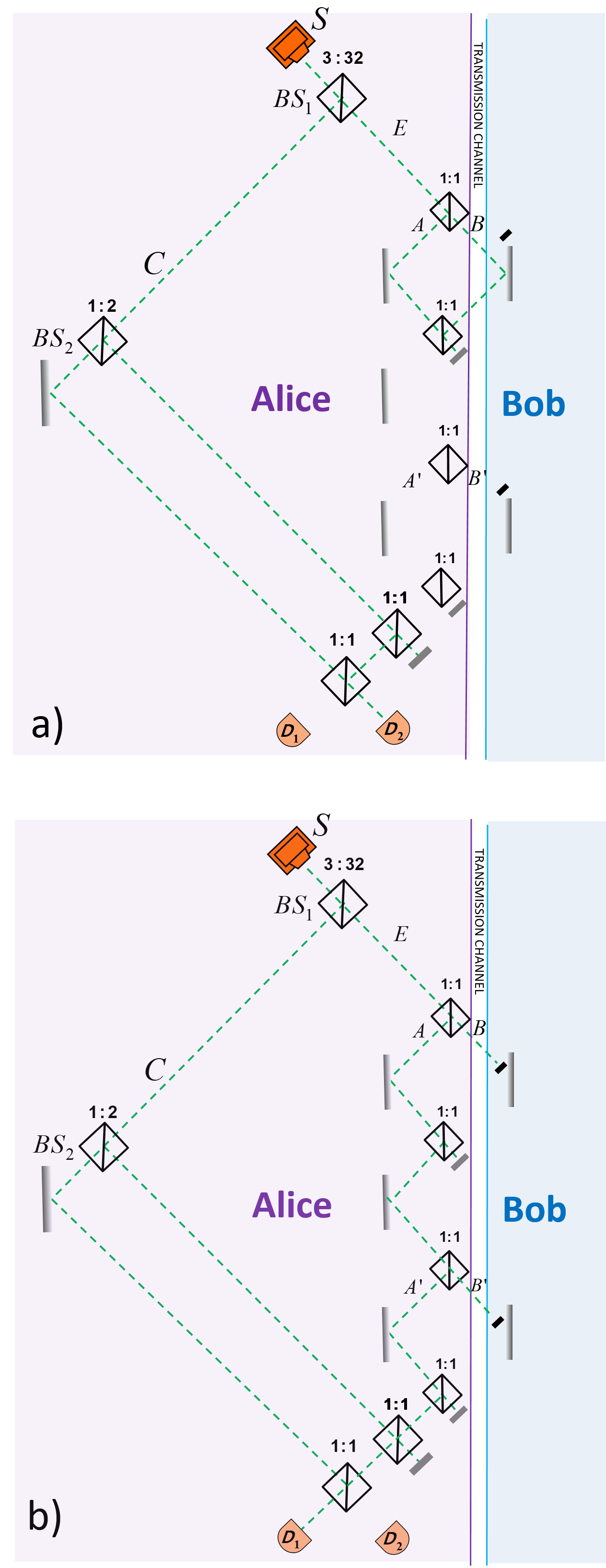}
     \end{center}
   \caption[example]
   { \label{fig:4}
Modified interferometer. a) It is tuned to
destructive interference in all internal MZIs toward left output ports and, therefore, to the destructive interference at detector $D_1$. b) It is also tuned such that  when arms $B$ and $B'$ are blocked, there is a destructive interference at detector $D_2$. }
   \end{figure}

The new protocol with ideal devices has zero error rate as the previous one, but now we  can also claim that no trace is left outside Alice's site. In Fig.~5a,b we describe the forward and backward wave functions in this protocol. The overlap of  provides the trace shown in Fig. 5c,d. There is no trace outside Alice's site for both bit values of the communicated bit.

 \begin{figure*} [h!]
   \begin{center}
     \includegraphics[height=17cm]{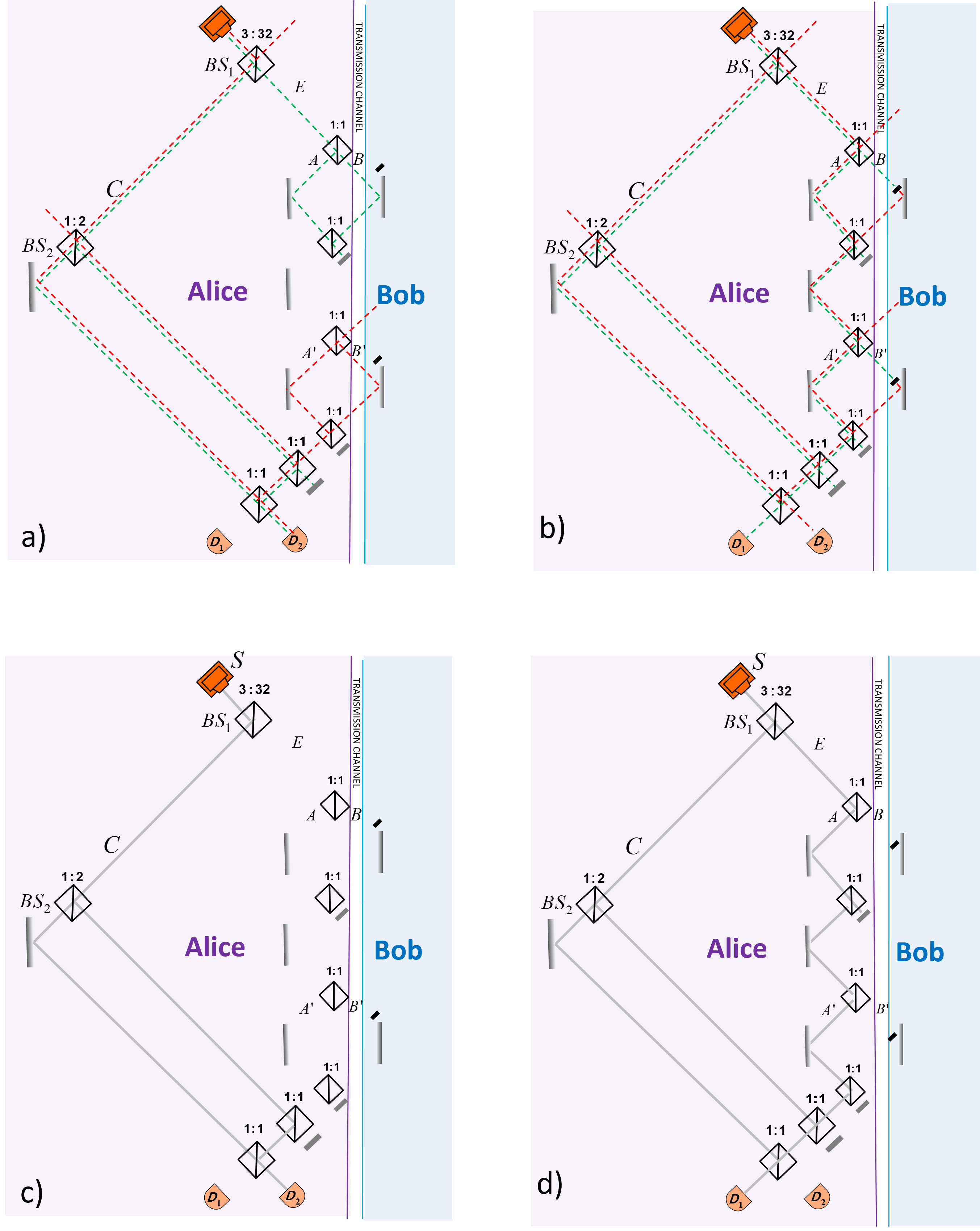}
     \end{center}
   \caption[example]
   { \label{fig:2}
Derivation of the trace left by the photon in the modified protocol using the two-state vector formalism. a) Forward and backward evolving waves of the photon for transfer of bit 0. b) Forward and backward evolving waves of the photon for transfer of bit 1. c) Trace of the photon for transfer of bit 0. d) Trace of the photon for transfer of bit 1. }
   \end{figure*}

To make it clear what do I mean by ``trace'' the photon leaves, let us consider a model \cite{PNAS} in which  the state of the photon passing through a channel is not changed, but the quantum state of each channel, originally described by  $| \chi \rangle$, is modified due to the passage  of the photon:
\begin{equation}\label{eq::iaSingle}
|\chi \rangle \rightarrow|\chi^\prime \rangle \equiv \eta |\chi \rangle + \epsilon |\chi^\perp \rangle ,
\end{equation}
where $| \chi^\perp \rangle$ denotes the component of $| \chi^\prime \rangle$ which is orthogonal to $| \chi \rangle$ and its phase is chosen such that $\epsilon > 0$. We assume that  $\epsilon \ll 1$.

In the protocol of Section \ref{sec:proto} for bit value 1 there is no trace outside Alice's site. The only place where the trace might be is arm $B$. The orthogonal component $|\chi^\perp \rangle{\!}_{_{B}}$ does appear during the evolution, but it is created  entangled with the spatial mode of the photon which is absorbed in the shutter Thus, the  component $|\chi^\perp \rangle{\!}_{_{B}}$ does not present in the branch with detection of the photon by Alice.

In the modified protocol for bit value 1, the situation is similar. The orthogonal components $|\chi^\perp \rangle{\!}_{_{B}}$ and $|\chi^\perp \rangle{\!}_{_{B'}}$ appear entangled with the photon modes which do not reach Alice's detectors.

After the click of Alice's detector,
there is a trace in arm $C$, as if a single photon passed there, the environment has the component
$\epsilon |\chi^\perp \rangle{\!}_{_{C}}\prod_{X\neq C}|\chi \rangle{\!}_{_X}$. Some other arms in Alice's site also have orthogonal components with first order in $\epsilon$ amplitude, but outside Alice's site there are no arms with orthogonal components.

For bit value 0, however, in the protocol of Section~\ref{sec:proto},  the trace does appear in arm $B$ outside Alice's site. The trace is the same as if we had a localized photon in arm $B$: $\epsilon |\chi^\perp \rangle{\!}_{_{B}}\prod_{X\neq B}|\chi \rangle{\!}_{_X}$. We also get the trace in arm $C$: $\epsilon |\chi^\perp \rangle{\!}_{_{C}}\prod_{X\neq C}|\chi \rangle{\!}_{_X}$. We cannot see simultaneously the traces in $B$ and in $C$, the traces are there, but they are entangled in such a way that when we detect one, the other disappear.

The interesting case is the bit value 0 in the modified protocol. The first order trace appears only in arms of the interferometer in Alice's site. The orthogonal components in arms $B$ and $B'$ appear only together with orthogonal component in an arm of another MZI, otherwise the photon mode on Bob's site cannot reach Alice's detector \cite{AV19}. The term in the quantum state of the environment with two orthogonal components has factor $\epsilon^2$. In the limit of better and better interferometer, when $\epsilon \rightarrow 0$, the trace is infinitely smaller than the trace left by a localized photon, and thus it is natural to neglect it and take into account only the first order trace. Note, that making modification with $N$ consecutive MZIs, instead of the two,  will lead to reducing the trace outside Alice's site to $N$-th order in $\epsilon$.

The modification by replacement of Bob's MZI by two (or $N$) consecutive MZIs can be applied to the  counterfactual communication protocol    which has high probability of success by applying Zeno effect \cite{Salih}. Even the counterfactual communication of a quantum state \cite{salih16,Li15,V16C} can be made  without first order trace outside  Alice's site.

\section{``Secondary'' trace }
\label{sec:second}

 For bit value 1 the original and modified protocols have no any trace outside Alice's site.   Even if the  interferometer is  not   ideal, in particular, when localized photon leaves some trace  passing through an arm of the interferometer, the photon in the protocol will not leave a trace outside Alice's site. The only requirement for full counterfactuality, i.e., no trace of any order, is that the shutter is $100\%$ opaque. Other imperfections might lead to errors, but cannot spoil counterfactuality.

 The situation for bit value 0 is different. The original protocol leaves a trace in Bob's site as if a localized photon was there. Even in the modified protocol there is some trace in Bob's site. Maybe a strongest argument that even the modified protocol is not counterfactual is that at least one of the nondemolition measurements of presence of the photon, one in arm $B$ and another in arm $B'$ performed together, will find the photon with certainty.

 I, however, view this correct statement as  a strong argument against considering counterfactual strong measurements for discussing presence of a pre and postselected quantum particle. Strong measurements change the situation, and properties which are true for a system with measurements might not be true when the measurements are not preformed.
 But there is also a trace argument.
 It is true that the local trace in Bob's site, i.e., appearing the orthogonal component  $  |\chi^\perp \rangle{\!}_{_{B}} $, is, at the limit of small coupling, is negligible. Not because it is just small, but because it is much smaller than the (small) trace the localized photon would leave there. Local trace is appearing of an orthogonal component of a quantum state of the environment in one location. However, the nonlocal trace - appearing orthogonal components in two locations, such as $  |\chi^\perp \rangle{\!}_{_{B}}   |\chi^\perp \rangle{\!}_{_{B^\prime}} $ - is the same as for the case of a localized wave packet of a single particle passing through the arms $B$ and $B'$. In both cases the trace is of second order in $\epsilon$. I named such a situation ``secondary presence'' \cite{secondary}.

 It is important to be aware of this secondary presence, as it provides an operational meaning for the difference between the transmission channel of the protocol which we consider counterfactual and the transmission channel in which no any communication takes place. However, it is natural to neglect nonlocal trace, not only because it is  of order $\epsilon^2$  and thus it is arbitrarily smaller than the local trace which is of order $\epsilon$, but also because there are no nonlocal interactions in Nature and observing of the effect of nonlocal trace requires special arrangement  of couplings  at different space time points to an external system.

 Is there a protocol in which for both bit values there is no {\it any} trace, not even a secondary trace in Bob's site?  It seems that  in a recent preprint \cite{SalRar} there is a claim of existence of such a protocol. An interferometric scheme employing manipulation of polarization of the photons allows to transmit bit value 0 without any trace on Bob's site even if the localized photon does leave a trace in the arms of the interferometer. However, I do not consider it as a counterexample. It is true, that  given a click at the detector  signifying bit 0, there is no any trace on Bob's site. However, for bit 0, the imperfection of the interferometer leads to a non-vanishing probability of a click of another Alice's detector which is a legitimate event in the protocol announcing bit 1. Together with the error, we also spoil the counterfactuality: in such an event, there is a trace on Bob's site.

\section{Discussion and conclusions}
\label{sec:discuss}

Counterfactual communication protocols are very paradoxical  phenomena, and the question: What is ``truly'' counterfactual?  remains controversial.   Arvidsson-Shukur,  Gottfries, and Barnes \cite{AGB}  applied sophisticated information tools (Fisher information, etc.) to evaluate counterfactuality of various protocols showing an advantage of protocols such as counterfactual key distribution \cite{Noh} in which there is no any  trace on Bob's site since only communication of bit 1 is used. It will be of interest to see the evaluation of the modified counterfactual  protocol  without first order weak local trace presented here.

Arvidsson-Shukur and  Barnes \cite{Arvid} proposed their own meaning of ``counterfactual'' communication: when particles go from Alice to Bob, while the information goes from Bob to Alice.  Indeed, it is easy to achieve such situation using the IFM scheme with Zeno effect \cite{Kwiat}. The scheme is counterfactual when Bob places the shutter: the particle detected by Alice  never reaches Bob and therefore, it does not go from Bob to Alice. On the other hand, when shutter is not placed, the photon does not come back to Alice, it ends up at Bob's site, so again, the particle does not go from Bob to Alice. I, however,  hardly see in this protocol justification  for the name ``counterfactual''. The photon was traveling between  Alice and Bob.  According to Arvidsson-Shukur and  Barnes definition of Alice's and Bob's sites, the photon goes to the transmission channel from Alice and comes back. If we  enlarge Alice's site to include the beam splitters, then the photon  does not go from the transmission channel to Alice, but then it goes from Bob to the transmission channel. We already discussed cases when the particle did not go the whole way between Alice and Bob.

The possibility of counterfactual communication without first order weak trace outside Alice's site and communication of bit 1 without any trace outside Alice's site is extremely paradoxical phenomena which  go  against the   spirit of science which searches for local causal explanation of Nature. It sounds like an action at a distance. My way to resolve the paradox \cite{PSA} is to accept the many-worlds  interpretation \cite{myMWI} which removes action at a distance on the level of all worlds together, explaining an illusion of action at a distance in our world.

To summarize, I proposed first counterfactual communication protocol which theoretically has no errors. The first version is counterfactual only if one considers traveling between Alice and Bob as following the whole continuous trajectory between them. The modified version is counterfactual based on  much stricter definition of having no first order trace outside Alice's site.  These protocols allowed to analyze various aspects of counterfactuality of  other proposals for counterfactual communication.

This work has been supported in part by the Israel Science Foundation Grant No. 1311/14.


\begin{thebibliography}{99}


\bibitem{Penrose}
R. Penrose,  {\em Shadows of the Mind}.
 Oxford: Oxford University Press  (1994).

\bibitem{IFM}
A. C. Elitzur,  and L. Vaidman,
 Quantum mechanical interaction-free measurements,
   Found.  Phys. {\bf 23}, 987 (1993).



\bibitem{Noh}
T.-G. Noh,
 Counterfactual quantum cryptography,
Phys. Rev. Lett. {\bf 103}, 230501 (2009).


\bibitem{Joz}
R. Jozsa,
Quantum effects in algorithms,
in {\it Lecture Notes in Computer Science},
C. P. Williams, ed. (Springer, London, 1998), Vol. 1509, p. 103.

\bibitem{Ho06}
O. Hosten, M.T.  Rakher, J.T. Barreiro, N.A. Peters, and P.G. Kwiat,
Counterfactual quantum computation through quantum interrogation,
Nature (London) {\bf 439}, 949 (2006).



\bibitem{Salih}
H. Salih, Z.H. Li, M. Al-Amri, and M.S. Zubairy,
Protocol for direct counterfactual quantum communication,
Phys.  Rev. Lett. {\bf 110}, 170502  (2013).



\bibitem{Wheeler}
J. A. Wheeler, The `Past' and the `Delayed-Choice Double-Slit Experiment', in Mathematical Foundations of Quantum Theory,
edited by A. R. Marlow (Academic Press, New York, 1978),
p. 9-48.

\bibitem{Berge}
B.G. Englert, K. Horia, J. Dai, Y.L. Len and H.K. Ng., Past of a quantum particle revisited. Phys. Rev. A {\bf 96}, 022126 (2017).



\bibitem{PV}
U. Peleg and L. Vaidman,
Comment on `Past of a quantum particle revisited', Phys. Rev. A, to be published, (2019).

\bibitem{PVrep}
B.G. Englert, K. Horia, J. Dai, Y.L. Len and H.K. Ng., Reply to ``Comment on `Past of a quantum particle revisited'  '', Phys. Rev. A, to be published, (2019).



\bibitem{past}
L. Vaidman,
Past of a quantum particle,
Phys.  Rev. A {\bf  87}, 052104 (2013).


\bibitem{AV90}
Y. Aharonov  and L. Vaidman,
  Properties of a quantum system during the time interval between two measurements,
  Phys. Rev. A  \textbf{41}, 11 (1990).


\bibitem{AV91}
Y. Aharonov  and L. Vaidman,
 Complete description of a quantum system at a given time,
J. Phys. A: Math. Gen. \textbf{24}, 2315 (1991).

\bibitem{Pusey}
I heard this comment from Matthew Pusey few years ago.


\bibitem{Gisin}
N. Gisin,
  Optical communication without photons,
  Phys. Rev. A  \textbf{88}, 030301 (2013).

\bibitem{Danan}
A. Danan, D. Farfurnik, S. Bar-Ad, and L. Vaidman,
Asking photons where they have been,
 Phys. Rev. Lett. \textbf{111}, 240402 (2013).


\bibitem{Shen13}
A. Shenoy, R. Srikanth, T. Srinivas,
Semi-counterfactual cryptography,
 Europhys. Lett. {\bf 103}, 60008 (2013).


\bibitem{Shen14}
A. Shenoy, R. Srikanth, T. Srinivas,
Counterfactual quantum certificate authorization,
 Phys. Rev. A  {\bf 89}, 052307 (2014).


\bibitem{Li14inv}
Z.-H. Li, M. Al-Amri, and M.S. Zubairy,
Direct quantum communication with almost invisible photons,
Phys. Rev. A {\bf 89}, 052334 (2014).

\bibitem{Guo14ent}
Q. Guo, L.-Y. Cheng, L. Chen, H.-F. Wang, and S. Zhang,
Counterfactual entanglement distribution without transmitting any particles,
Opt. Express {\bf 22}, 8970 (2014).


\bibitem{guo14}
Q. Guo, L.-Y. Cheng, L. Chen, H.-F. Wang, and S. Zhang,
Counterfactual quantum-information transfer without transmitting
any physical particles,  Phys. Rev. A  {\bf 90}, 042327 (2014).



\bibitem{guo15}
Q. Guo, L.-Y. Cheng, L. Chen, H.-F. Wang, and S. Zhang,
Counterfactual quantum-information transfer without transmitting
any physical particles, Sci. Rep. {\bf 5}, 8416 (2015).


\bibitem{Li15}
Z.H. Li, M. Al-Amri, and M.S. Zubairy,
Direct counterfactual transmission of a quantum state,
Phys. Rev. A {\bf 92}, 052315 (2015).


\bibitem{chen15}
Y.Chen, X. Gu, D. Jiang, L. Xie, and L. Chen,
Tripartite counterfactual entanglement distribution,
Opt. Express {\bf 23} 21193 (2015).



\bibitem{Shen15}
A. Shenoy, and R. Srikanth,
Counterfactual distribution of {S}chr\"odinger cat states,
 Phys. Rev. A  {\bf 92}, 062308 (2015).

\bibitem{chen16}
Y.Chen, D. Jiang, X. Gu, L. Xie, and L. Chen,
Counterfactual entanglement distribution using quantum dot spins,
J. Opt. Soc. Am. B {\bf 33} 663  (2016).


\bibitem{salih16}
H. Salih,  Protocol for counterfactually transporting an unknown qubit, Front. Phys.  {\bf 3}, 94 (2016), arXiv:1404.2200 (2014).


\bibitem{Guo17}
Q. Guo, L.-Y. Cheng, L. Chen, H.-F. Wang, and S. Zhang,
Counterfactual quantum-information transfer without transmitting
any physical particles, Phys. Rev. A {\bf 96}, 052335 (2017).


\bibitem{cao17}
Y. Cao, Y.-H. Li, Z. Cao, et al.
Direct counterfactual communication via quantum Zeno effect,
Y. Cao, Y.-H. Li, Z. Cao, J. Yin, Y.-A. Chen, H.-L. Yin, T.-Y. Chen, X. Ma, C.-Z. Peng, and J.-W. Pan,
Proc. Natl. Acad. Sci. USA {\bf 114},    4920 (2017).

\bibitem{liu17}
C. Liu, J. Liu, J. Zhang, and S. Zhu,
The Experimental demonstration of high efficiency interaction-free measurement for quantum counterfactual-like communication,
Sci. Rep. {\bf 7},  10875  (2017).


\bibitem{zaman18}
F. Zaman,  Y. Jeong, L. Chen,   and H. Shin,
Counterfactual quantum-information transfer without transmitting
any physical particles, Sci. Rep. {\bf 8}, 14641 (2018).


\bibitem{guo18}
Q. Guo, L.-Y. Cheng, H.-F. Wang, and S. Zhang,
Counterfactual entanglement swapping enables high-efficiency entanglement distribution,
Opt. Expr. {\bf  26},  27314  (2018).


\bibitem{zhang13}
J.-L. Zhang, F.-Z. Guo, F. Gao, B. Liu, and Q.-Y. Wen,
Private database queries based on counterfactual quantum key distribution,
Phys. Rev. A {\bf 88}, 022334 (2013).


\bibitem{salih14tri}
H. Salih,
Tripartite counterfactual quantum cryptography,
Phys.  Rev. A {\bf 90}, 012333  (2014).

\bibitem{K-Du}
F. Kong, C. Ju, P. Huang, P. Wang, X. Kong, F. Shi, L. Jiang, and J. Du,
Experimental realization of high-efficiency counterfactual computation,
 Phys. Rev. Lett. {\bf 115}, 080501 (2015).


\bibitem{V07}
L. Vaidman,
 Impossibility of the counterfactual computation for all possible outcomes,
Phys. Rev. Lett. {\bf 98}, 160403 (2007).


\bibitem{V14}
L. Vaidman,
 Comment on ``Protocol for direct counterfactual quantum communication'',
Phys. Rev. Lett. {\bf 112}, 208901 (2014).

\bibitem{V14R}
H. Salih, Z.H. Li, M. Al-Amri, and M.S. Zubairy,
Salih et al. Reply,
Phys. Rev. Lett. {\bf 112}, 208902 (2014).


\bibitem{V15}
L. Vaidman,
Counterfactuality of ‘counterfactual’ communication,
J. Phys. A: Math. Theor. {\bf  48},  465303 (2015).


\bibitem{V16C}
L. Vaidman,
Comment on ``Direct counterfactual transmission of a quantum state''
Phys. Rev. A {\bf 93}, 066301 (2016).

\bibitem{V16R}
Z.-H. Li, M. Al-Amri, and M. S. Zubairy,
Reply to ``Comment on ‘Direct counterfactual transmission of a quantum state’”,
Phys. Rev. A {\bf 93}, 066302 (2016).

\bibitem{AV19}
Y. Aharonov  and L. Vaidman,
  Properties of a quantum system during the time interval between two measurements,
  Phys. Rev. A , to be published,  (2019).

\bibitem{PNAS}
J. Dziewior, L. Knips, D. Farfurnik, K. Senkalla, N. Benshalom, J. Efroni, J. Meinecke, S. Bar-Ad, H. Weinfurter, and L. Vaidman,
 Universality property of local weak interactions and its application for interferometric alignment, Proc. Natl. Acad. Sci. USA, to be published,
 arXiv:1804.05400 (2018).


\bibitem{secondary}
L. Vaidman,
Tracing the past of a quantum particle,
Phys. Rev. A {\bf 89}, 024102 (2014).

\bibitem{CFInt5}
X. Liu, B. Zhang, J. Wang, C. Tang, J. Zhao, and S. Zhang,
Eavesdropping on counterfactual quantum key distribution with finite resources,
Phys. Rev. A {\bf 90}, 022318  (2014).

\bibitem{SalRar}
H. Salih, W. McCutcheon, J. Hance, P. Skrzypczyk, and J. Rarity,
Do the laws of physics prohibit counterfactual communication?
arXiv:1806.01257 (2018).



\bibitem{AGB}
D.R.M. Arvidsson-Shukur, A.N.O. Gottfries, and C.H.W. Barnes,
Evaluation of Counterfactuality in Counterfactual Communication Protocols,
Phys. Rev. A {\bf 96}, 062316 (2017).


\bibitem{Arvid}
D. R. M. Arvidsson-Shukur and C. H. W. Barnes
Phys. Rev. A 94,
Quantum counterfactual communication without a weak traces,
Phys. Rev. A {\bf 94}, 062303 (2016).


\bibitem{Kwiat}
P. Kwiat, H. Weinfurter, T. Herzog, A. Zeilinger, and M. Kasevich,
Interaction-Free Measurement,
Phys. Rev. Lett. {\bf 74}, 4763 (1995).



\bibitem{PSA}
L. Vaidman,
 On the paradoxical aspects of new quantum experiments, in [PSA: Proceedings of the Biennial Meeting of the Philosophy of Science Association 1994]  211, Philosophy of Science Association (1994).



\bibitem{myMWI}
 L.~Vaidman,  Many-Worlds Interpretation of Quantum
Mechanics, {\it Stan. Enc. Phil.},  E. N. Zalta (ed.) (2002),
http://plato.stanford.edu/entries/qm-manyworlds/.







\end{thebibliography}
\end{document}